\documentclass[11pt,a4paper]{article}
\usepackage{jheppub}
\usepackage{longtable}
\usepackage{bm}

\usepackage{graphicx}
\usepackage{dcolumn}
\usepackage{bm}
\usepackage{amssymb}

\newcommand*{\La}{\cal{L}}
\newcommand*{\no}{\noindent}
\newcommand*{\bea}{\begin{eqnarray}}
\newcommand*{\eea}{\end{eqnarray}}
\newcommand*{\be}{\begin{equation}}
\newcommand*{\ee}{\end{equation}}
\newcommand*{\pd}{\partial}
\newcommand*{\pdm}{\pd_{\mu}}

\newcommand*{\pref}[1]{(\ref{#1})}

\newcommand*{\mn}{{\mu\nu}}

\newcommand*{\nn}{\nonumber}

\title{(Non-)Aligned gauges and global gauge symmetry breaking}

\author[1]{Axel Maas}\emailAdd{axelmaas@web.de}

\affiliation[1]{Institute for Theoretical Physics, Friedrich-Schiller-University Jena, Max-Wien-Platz 1, D-07743 Jena, Germany}

\abstract{The concept of (global) gauge symmetry breaking plays an important role in many areas of physics. Since the corresponding symmetry is a gauge symmetry, its breaking is actually gauge-dependent. Thus, it is possible to design gauges which restore the symmetry as good as possible. Such gauge constructions will be detailed here, illustrated with the use of lattice gauge theory. Their use will be discussed for the cases of the Higgs effect, high-baryon density color superconductors, and BRST symmetry.} 

\keywords{Global gauge symmetry, gauge fixing, Higgs effect, color superconductors, BRST symmetry (PACS 11.15.Ex)}

\begin{document}

\maketitle

\section{Introduction}

A symmetry of a classical system, i.\ e.\ of the Lagrangian, which remains unbroken in the quantization process can still be broken dynamically, i.\ e.\ spontaneously. To describe this, in the following the language of correlation functions will be used, which is sufficient, since these encode all information of a theory \cite{Bohm:2001yx}. In terms of correlation functions, a symmetry is unbroken, if under a symmetry transformation of the Lagrangian for fields $\phi_i$
\be
\phi_i \to\phi_i+ \delta\phi_i \label{st}
\ee
\no all correlation functions remain invariant as well. This implies that the correlation functions of operators not invariant under \pref{st}, e.\ g.\ $\langle\phi_i\rangle$, have to vanish identical. If these conditions are not fulfilled, the symmetry is broken.

In case of a gauge theory, it is necessary to differentiate between gauge-invariant and non-gauge-invariant correlation functions. If only gauge-de\-pen\-dent correlation functions show a behavior indicating the breakdown of a symmetry, this cannot have any measurable consequences \cite{Fradkin:1978dv,Caudy:2007sf,Greensite:2008ss}. If the affected symmetry is a global part of the gauge symmetry\footnote{Any local part cannot be broken anyhow \cite{Elitzur:1975im}.}, it can be argued that this is a mere artifact of the description, and the symmetry is, in fact, intact \cite{Lenz:1994tb,Perez:2008fv,Lenz:2000zt,Frohlich:1981yi,Frohlich:1980gj}. This is the combination of observations which will be used here.

Before continuing, it is worthwhile to discuss the question of observing a broken symmetry. Naively, when just performing the path integral, all correlation functions non-invariant under some global symmetry will always vanish\footnote{The fact that that does not seem to be the case in numerical lattice simulations is just an artifact of the employed local algorithms, which fail to perform the full average over field configuration space in this case. The problem is that when starting in a certain sector of the theory, the update cycles often stay within this starting sector.}. The trivial reason is just that no direction of the global symmetry is preferred without an explicit breaking. This can only be remedied by introducing an explicit breaking when determining the correlation functions, and performing the limit of zero explicit breaking afterwards. This is done, e.\ g., by the introduction of an external field.

In fact, in an experimental observation an absolute direction is never observed. E.\ g., take a magnet's magnetization. The orientation is measured relative to a reference orientation, e.\ g., compared to a different magnet. This reference orientation plays the role of the explicit breaking. Thus, an equivalent way of observing the breaking of a symmetry is to use operators invariant under a global symmetry transformation, but measuring the relative orientation of two operators not being separately invariant. One such operator for a magnet would be the averaged magnetization correlator,
\bea
C _{MM}&=& \langle MM \rangle \label{mtest} \\
M&=&\sum_i s_i \nn
\eea
\no where the $s_i$ are the local spins, and thus $M$ is the configuration-wise polarization. Since the latter is only non-zero in the broken phase, the correlator can only then be non-zero.

In a gauge theory, the explicit breaking of the global gauge symmetry is introduced by a gauge condition \cite{Frohlich:1981yi,Frohlich:1980gj}, e.\ g.\ the 't Hooft condition in a gauge-Higgs systems \cite{Bohm:2001yx}. The gauge condition is itself not invariant under a global gauge transformation, and the operators signaling the breaking of the symmetry have the same type of non-invariance. E.\ g., take the 't Hooft gauge condition
\be
\pdm W^\mu_a+i\xi g\tau^a_{ij}\langle\phi_i\rangle(\phi_j-\langle\phi_j\rangle)=0\label{thooft}
\ee
\no involving the gauge field $W_\mu$, the gauge coupling $g$, the gauge parameter $\xi$, the Higgs field $\phi$ and its expectation value $\langle\phi_i\rangle$. The latter is chosen in a particular direction $n_i$ in color space, $\langle\phi_i\rangle=v n_i$, thereby explicitly implementing a breaking of the global gauge invariance, like an external field. This choice, or even the use of 't Hooft gauge, no matter how technically useful it is, is arbitrary. Thus, it is more precise to speak of a hidden, rather than a broken, symmetry \cite{Bohm:2001yx}. The continued existence of the symmetry manifests itself in the presence of Ward-Takahashi and Slavnov-Taylor identities (STIs), relating quantities which are independent of each other in a theory with explicit breaking.

The aim here is to use this knowledge to construct gauges which hide as little symmetry as possible, and which will be aptly called non-aligned gauges. They will be compared to gauges preferring a global direction, like the 't Hooft gauge. Such gauges will be called aligned gauges. As the paradigmatic example the gauge-Higgs system will be discussed in section \ref{higgs}. Further examples will be the recovery of BRST symmetry beyond perturbation theory in Yang-Mills theory in section \ref{brst}, and color superconductors in section \ref{density}. A brief concluding remark is given in section \ref{summary}. The main purpose of the present exercise is to construct gauges making use as much as possible of the explicit symmetry. This facilitates the determination of gauge-dependent correlation functions by having as few non-zero correlation functions as possible, thus having simpler relations between connected and disconnected correlation functions, and simpler color tensor structures, see section \ref{higgs}. 

\section{Higgs effect}\label{higgs}

The simplest, and at the same time most paradigmatic, example is a fundamental complex scalar $\phi$ coupled to a SU(2) valued gauge fields $W_\mu$,
\be
{\La}=-\frac{1}{4}W_\mn^aW^\mn_a+(D_\mu\phi)^+D^\mu\phi-\gamma(\phi\phi^+)^2-\frac{v}{2}\phi\phi^+\label{la},
\ee
\no with the field strength tensor $W_\mn$ and the covariant derivative $D_\mu$, both containing the gauge coupling $g$. To construct the desired gauge, it is necessary to fix the local gauge degree of freedom. For this, minimal Landau gauge is chosen here, though of course any other non-perturbative resolution of the Gribov-Singer ambiguity can equally well be chosen \cite{Maas:2011se}\footnote{Note that it can be argued that the Gribov horizons in the Higgs phase should become irrelevant \cite{Lenz:2000zt}, thus making this choice less relevant for the gauge-dependent correlation functions. This is supported by lattice simulations \cite{Maas:2010nc}. Note that the Higgs phase is not separated from the confinement phase \cite{Fradkin:1978dv,Caudy:2007sf}. They should be denoted as a would-be Higgs phase and a would-be confinement phase, though for the sake of brevity the would-be will be suppressed here.}. Furthermore, the following applies equally to any covariant gauge, Coulomb gauge, or any other gauge, which is not referring explicitly to the global degree of gauge freedom, i.\ e., has a gauge condition which is invariant under global gauge transformations. Thus, 't Hooft gauges or maximal Abelian gauges are explicitly excluded.

All of these gauges leave the global gauge symmetry invariant, which is irrelevant without any breaking of the symmetry, but yields an unspecified behavior in case of breaking. To remedy the situation, an additional gauge condition for the global part of the gauge fixing is required. To fix this, a specification using the correlation functions will be used here, as can already be done for the local part of the gauge condition \cite{Maas:2011se}. Using the Higgs field $\phi_i$, the two choices investigated here will be aligned supplementary gauge conditions
\be
\langle\phi_i\rangle=vn_i \label{agc}
\ee
\no and non-aligned supplementary gauge conditions
\be
\langle\phi_i\rangle = 0 \label{nagc},
\ee
\no where $n_i$ is an arbitrary, complex, two-dimensional unit vector. The number $v$ cannot be determined by the gauge condition alone. It is the Higgs vacuum expectation value, and is determined by the dynamics. 

The aligned gauges are essentially the Landau-gauge limit $\xi\to 0$ of the 't Hooft gauge \pref{thooft}. However, the non-aligned gauge will not exhibit a Higgs condensation in the conventional way. Of course, this implies that the conventional perturbative expansion around the classical Higgs mean field in a semi-classical way \cite{Bohm:2001yx} will not be possible, and most of the Higgs physics is therefore relegated to pure non-perturbative quantum effects, making a non-perturbative treatment mandatory.

The operational definition of such gauge conditions is straightforward in both lattice and continuum calculations. In continuum calculations, just the conditions \pref{agc} and \pref{nagc} have to be implemented. One possible realization is as constraints on the solutions, e.g. of functional equations. Alternatively, on a more fundamental level in the path integral this can be performed either by a limit process of the 't Hooft gauges for aligned gauges, or by performing the path integral unrestricted over the global gauge degree of freedom for the non-aligned case. For a lattice implementation, just a global gauge transformation after local gauge fixing is necessary to bring a configuration into an aligned gauge. For the non-aligned gauge, either an average over many gauge copies with different global gauge transformations can be performed, or, because of the absence of a global weight factor, just one random representative can be chosen. The latter approach takes advantage of the ergodicity of a lattice algorithm in a fixed sector at the price of larger statistical fluctuation around the non-aligned values.

Of course, neither gauge choice affects quantities measuring relative orientations, e.g. the quantity \cite{Caudy:2007sf}
\bea
\Phi=\langle\phi^+\phi\rangle\label{rel} \\
\phi= \int d ^dx \phi(x)\nn
\eea
\no still remains a measure of whether a preferred alignment of the Higgs in a given configuration exists, or not. It is the generalization to the global gauge symmetry of the magnetization test \pref{mtest}. It should be noted that \pref{rel}, and also \pref{mtest}, will only in an infinite volume mark the state of the symmetry \cite{Caudy:2007sf}. In any finite volume, due to finite volume artifacts, they will remain always non-zero. Thus, in lattice simulations their infinite-volume extrapolation will be required.

\begin{figure}
\includegraphics[width=\linewidth]{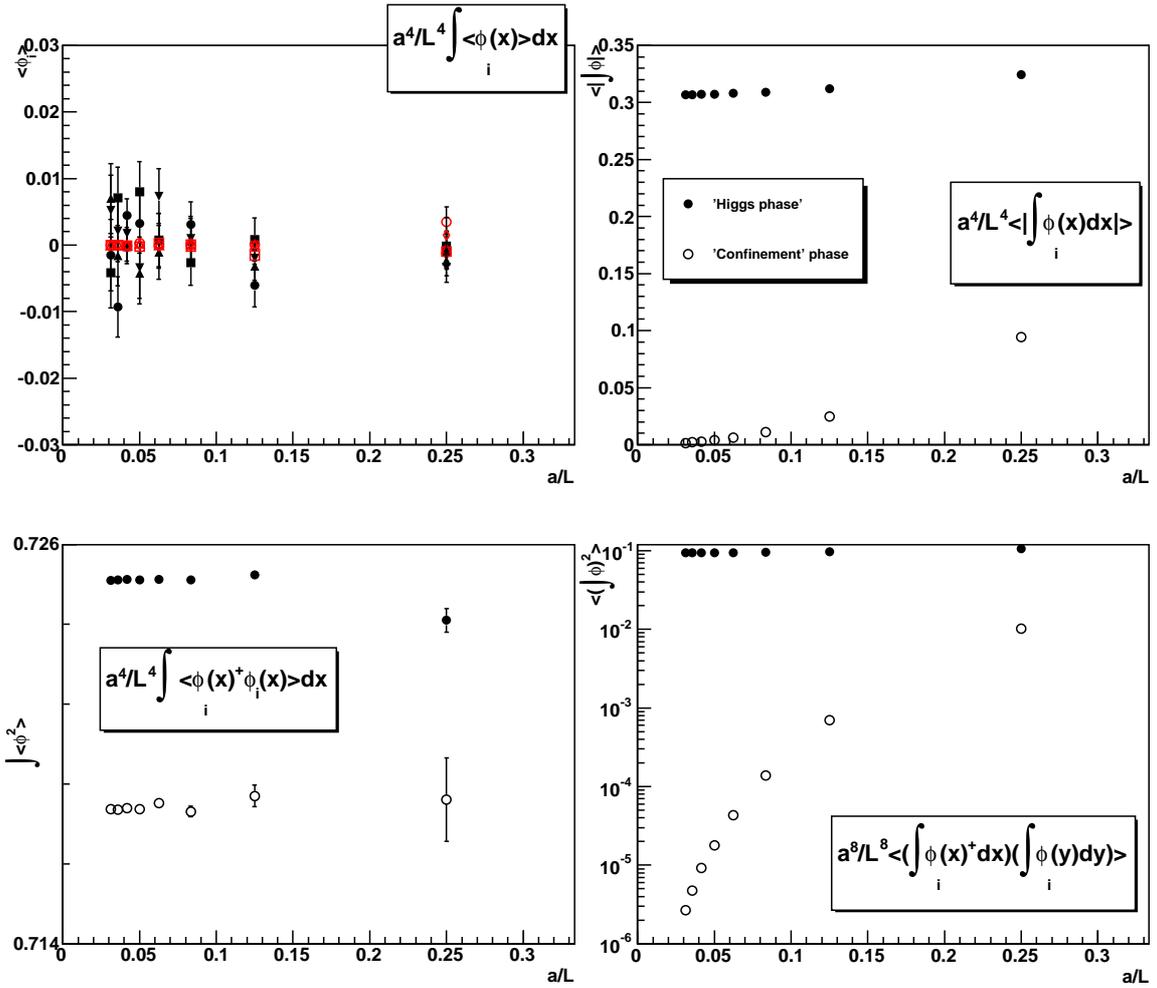}\\
\caption{\label{fig}Results for both the Higgs phase and the confinement phase in a non-aligned gauge, as a function of the lattice extent $L$ in units of the lattice spacing $a$. All results have been obtained using lattice simulations with lattice parameters $\beta=2.3$, $\kappa=0.32$, $\lambda=1.0$ for the Higgs phase and $\beta=2.0$, $\kappa=0.25$, $\lambda=0.5$ for the confinement phase, have not been renormalized, and are given in lattice units. See \cite{Maas:2010nc} for details of the simulations. For the continuum Lagrangian \pref{la}, these parameters translate to the values $g=1.3$, $\gamma=9.8$, $v=-11$, and $g=1.4$, $\gamma=8$, and $v=-7.8$, respectively. Top-left panel: The Higgs expectation value. In aligned gauges, the component along the aligned direction would be non-zero. Top-right panel: Absolute value of the average Higgs value. It is clearly non-zero in the Higgs phase. Furthermore, this would be the size of the Higgs expectation value in the non-aligned direction. Bottom-left panel: Average fluctuation of the Higgs field. Non-zero in both phases. Bottom-right panel: Alignment test quantity \pref{rel}. It clearly distinguishes between the Higgs and the confinement phase. Nonetheless, though being invariant under global gauge transformations, it could change under local gauge transformations. Note that in units of the bound state masses the physical units are of the same order of magnitude in both cases \cite{Maas:2010nc}.}
\end{figure}

To illustrate these concepts, the various quantities are shown in figure \ref{fig}, using lattice gauge theory and performing the infinite-volume extrapolation. Thus, in the non-aligned gauges no manifest breaking of the global gauge symmetry remains. Of course, the dynamical difference between the would-be Higgs phase and the would-be confinement phase remains, as signaled by quantities like \pref{rel}.

This justifies to use even in the would-be Higgs phase a much simpler description in terms of color-diagonal quantities in a non-aligned gauge. E.\ g., in an aligned gauge the connected, indicated by a subscript $c$,  propagator of the Higgs field is given by
\bea
<\phi_i\phi_j>_c&=&<\phi_i\phi_j>-<n_i\phi_i>^2\nn\\
<\phi_i\phi_j>_c&=&n_i n_j D_L+(\delta_{ij}-n_i n_j) D_T\nn
\eea
\no with two dressing functions $D_L$ and $D_T$. For the non-aligned gauge the connected propagator is given by
\be
<\phi_i\phi_j>=<\phi_i\phi_j>_c=\delta_{ij}D\nn,
\ee
\no with just a single dressing function $D$. For vertices, the simplifications are even more substantial. Of course, since this is a pure gauge choice, no gauge-invariant quantities are affected by this choice. Especially, the continuous connection of the Higgs and confinement phase in terms of gauge-invariant physics \cite{Fradkin:1978dv} remains unaffected.

\section{BRST symmetry}\label{brst}

The same concept gives a new perspective on the construction of a global BRST symmetry \cite{Hirschfeld:1978yq,Neuberger:1986xz,Kalloniatis:2005if,vonSmekal:2008en,Mehta:2009zv} in Yang-Mills theory beyond perturbation theory in the continuum. In covariant gauges in perturbation theory, a BRST transformation of the gluon, the ghost and anti-ghost takes the form \cite{Bohm:2001yx}
\bea
\delta_\mathrm{BRST}A_\mu^a(x)&=&\delta\lambda D_\mu^{ab}c_b(x)\label{brst:trans1}\\
\delta_\mathrm{BRST}c^a(x)&=&-\delta\lambda\frac{1}{2}g_df^{abc}c_b(x)c_c(x)\\
\delta_\mathrm{BRST}\bar c^a(x)&=&\delta\lambda\frac{1}{\xi}\pdm A_\mu^a(x)\label{brst:trans3},
\eea
\no with Grassmann parameter $\delta$ and ghost and anti-ghost field $c$ and $\bar c$. The transformation is just an ordinary gauge transformation for the gluon fields. Especially, if the ghost fields are integrated out, the gauge function $\delta c_a(x)$ would become just a complicated, likely non-local, functional, depending on the gauge fields such that not only the action would be left invariant, but also the Faddeev-Popov determinant, possibly up to a weight factor. What happens is essentially just a gauge transformation between the different gauge copies satisfying the covariant gauge condition. This implies that the Gaussian weight function will change such as to alter the weight of the gauge copy appropriately. Furthermore, in the Landau gauge case the perturbative BRST transformation can only be the identity transformation, as there is one and only one configuration perturbatively which satisfies the Landau gauge condition.

Beyond perturbation theory, the situation changes by the appearance of Gribov copies \cite{Gribov:1977wm,Maas:2011se}. Especially, since there are now more than one gauge copy satisfying the Landau gauge condition, the BRST transformation is no longer an identity transformation in Landau gauge.

Now, concentrate on Landau gauge. By construction, any BRST transformation will not change the Landau gauge condition, because of the anti-ghost equation of motion. This implies that BRST transformations mediate between different Landau gauge Gribov copies. If an extended gauge condition, like minimal Landau gauge \cite{Maas:2011se}, selects only a subset of Gribov copies, it is possible that a BRST transformation leads out of this set, and thus BRST symmetry appears broken \cite{Maas:2011se,Dudal:2008sp}: The correlation functions are no longer invariant. However, this breaking is again introduced by an additional gauge condition, which specializes the Landau gauge further to the minimal Landau gauge.

To restore invariance, it is, as before, necessary to choose a gauge which is not aligned with respect to BRST. In fact, such a gauge has already been constructed, the Landau-Hirschfeld-Neuberger-von Smekal gauge \cite{Hirschfeld:1978yq,Neuberger:1986xz,Kalloniatis:2005if,vonSmekal:2008en,Mehta:2009zv}, which averages over all Gribov copies and has been shown to harbor an intact BRST symmetry\footnote{lt would of course be interesting to know if there exist any subsets of Gribov copies such that BRST would remain intact within these subsets. This would be very interesting from a practical point of view, but currently remains pure speculation \cite{Maas:2011se,Maas:2010wb}.}. Thus BRST symmetry can be regained just as any other global gauge symmetry. Whether there exist relative alignment operators and what their interpretation would be is an interesting question, in particular whether they show the existence of broken and unbroken phases, and if yes, under which circumstances.

\section{Color superconductors }\label{density}

The existence of exotic phases of matter at high baryon densities has long been conjectured \cite{Leupold:2011zz}. In various calculations, these phases show color superconductivity by the condensation of some colored states, e.g. in color octet channels. Such condensates play a role similar to the Higgs condensate in the gauged Higgs case before. In particular, their values depend on the gauge \cite{Rajagopal:2000wf}. However, due to the Lorentz structure and the composite nature of the condensates, it is not possible to write down a direct analogue of the 't Hooft gauges. This also implies that, in an appropriate non-aligned gauge, such colored condensates will vanish. Of course, the break-down of the global symmetry in the sense of relative alignments is still possible.

The fundamental problem here is to give an equally simple global gauge condition as for the Higgs case\footnote{A possibility may be to perform a Hubbard-Stratonovitch transformation \cite{Hubbard:1959ub,Stratonovitch:1957st} at the classical level, and then use the auxiliary scalar Hubbard-Stratonovitch field to formulate the gauge condition. I am grateful to Holger Gies for suggesting this possibility.}, though the operative definitions remain for the non-aligned gauge. However, now the vanishing of the colored condensates has to be demanded. An implementation of the aligned gauges will likely require a composite external source, a complication I will not delve into here. In functional calculations like those in \cite{Nickel:2006vf,Nickel:2006kc,Nickel:2008ef}, such an alignment is indeed performed implicitly, offering at least one operative implementation.

Finally, one remark must be made about the multitude of model calculations of QCD at high densities, in which the gauge symmetry is replaced by a global symmetry \cite{Leupold:2011zz,Buballa:2003qv}, thus making gauge choices in the conventional sense impossible. In these cases, colored condensates have a well-defined meaning. This implies that comparisons to full QCD have then to be performed with care. If objects with open color indices are compared, it is necessary that an appropriate aligned gauge is used for QCD. Furthermore, the investigation of the phase structure cannot be based on colored condensates alone, since any changes of them in full QCD may be gauge-dependent, even if they are not in the model. This has already been observed in a multitude of other theories \cite{Caudy:2007sf,Greensite:2008ss}. Fortunately, usually thermodynamic quantities also clearly signal this information in such models \cite{Leupold:2011zz,Buballa:2003qv}. This is not surprising, as the condensates in such theories have a physical reality, and thus should resemble the phase structure. This is no longer necessarily the case in QCD.

\section{Summary}\label{summary}

By the choice of an aligned or non-aligned gauge, it is possible to be either manifestly sensitive or insensitive to global phase changes. However, this does not affect anything only sensitive to relative global phases. Such gauges therefore offer the possibility to work with correlation functions of maximal symmetry, which simplifies, e.\ g. functional calculations, reduces the statistical requirements of lattice calculations, and simplifies STIs. They are therefore of great practical use for calculations using gauge-dependent correlation functions in intermediate states. It is also not necessary to include semi-classical considerations concerning possible condensate structures in theories with a Higgs-like effect to define the gauges. Furthermore, the concept of such gauges is helpful to design settings with particular desirable properties, e.g. global BRST.\\

{\bf Acknowledgments}\\

The author is grateful to Holger Gies, Jeff Greensite, Lorenz von Smekal, and Michael Buballa for very helpful discussions. This work was supported by the DFG under grant number MA 3935/5-1.

\bibliographystyle{bibstyle}
\bibliography{bib}


\end{document}